# Decoupling of $\delta^{13}C_{carb}$ and $\delta^{13}C_{org}$ during the Carnian Pluvial Episode

Enhao Jia[a], Kui Wu[b], Yong Du[a], Yuyang Wu[a], Fengyu Wang[a], Xu Dai[a], Huyue Song[a], Daoliang Chu[a], Lei Zhong[a], Zhiwei Yuan[a], Xiangmin Chen[a], Zhe Li[c], Haijun Song[a]*

[a] School of Earth Sciences, State Key Laboratory of Biogeology and Environmental Geology, China University of Geosciences, Wuhan 430074, China

[b] Hubei Institute of Geosciences, Hubei Key Laboratory of Resource and Ecological Environment Geology, Hubei Geological Bureau, Wuhan 430034, China

* Corresponding author's e-mail: haijunsong@cug.edu.cn

**Abstract**

The Carnian Pluvial Episode (CPE) was a major global climate change event in the early Late Triassic that significantly affected marine ecosystems and carbon cycles. One of the most prominent features of the CPE is the coupled multiple negative $\delta^{13}C_{carb}$-$\delta^{13}C_{org}$ excursions. However, at Erguan (Nanpanjiang Basin) and Xiashulao (Lijiang Basin) from eastern Tethys, a decoupling between $\delta^{13}C_{carb}$ and $\delta^{13}C_{org}$ during CPE was observed. At the end of early Carnian (Julian), the $\delta^{13}C_{carb}$ showed a negative excursion of 2-3‰, while the $\delta^{13}C_{org}$ exhibited a positive excursion of about 3-4‰. In addition, increased terrestrial inputs is indicated by the rising C/N (3 to 10) and decreasing Y/Ho (42 to 27) that coexist with this decoupling. Carbon isotope records from other regions provide further evidence that decoupling is unlikely to be caused by regional diagenetic processes. A correlation between the sedimentary environment and the presence of decoupling was found. The coupling of $\delta^{13}C$ negative excursions is from the shallow shelves and the deep slopes, whereas the decoupling occurs from the deep shelf to the shallow slope. In the deep shelf to the shallow slope, sedimentary organic matter is mainly sourced from pelagic before the CPE as evidenced by low C/N (~3) and high Y/Ho (~36-42). During the CPE, the increased fresh water flux (Sr/Ba <1) enhanced terrestrial input in organic matter, which may cause positive excursions in the $\delta^{13}C_{org}$ record with elevated TOC content. As a result, the $\delta^{13}C_{org}$ and $\delta^{13}C_{carb}$ decoupled. In contrast, organic matter in sediments from the shallow shelf and deep slope are mainly from terrestrial and pelagic sources, respectively. This study reveals the significant impact of terrestrial inputs on marine carbon cycling during the Carnian Pluvial Episode, highlighting the crucial role of climate events in modifying the carbon isotope record.

## 1. Introduction

The Carnian Pluvial Episode (CPE) was an important global climate change event in the early Late Triassic (Simms and Ruffell, 1989; Dal Corso et al., 2020). Its most notable feature was a significant intensification of the hydrological cycle, with multiple instances of increased rainfall recorded from Pangaea to the Tethys Ocean (Ruffell et al., 2016; Barrenechea et al., 2018; Dal Corso et al., 2020). In sedimentary records, this event is indicated by the widespread development of clastic lithofacies in several global Carnian sedimentary sequences, associated with increased terrestrial input due to pluvial events (Simms and Ruffell, 1989; Dal Corso et al., 2015). Major transformations in biological communities within both marine and terrestrial ecosystems coincided with the CPE, for instance, the emergence of skeletal coral reefs and calcareous nannofossils in marine environments (Dal Corso et al., 2020; Dal Corso et al., 2021), and the elevated extinction rates of conodonts (Zhang et al., 2018). On land, significant diversification of dinosaurs and the origin of mammals occurred (Bernardi et al., 2018; Lucas, 2018; Dal Corso et al., 2020).

The major global changes during the CPE have been shown to be associated with large-scale disturbances in the carbon cycle, characterized by multiple negative $\delta^{13}C$ excursions in marine carbonate and terrestrial organic carbon isotope records (Dal Corso et al., 2018; Lu et al., 2021; Zhang et al., 2022). However, the pattern of global carbon isotope disturbances during the Carnian is unclear. The number of $\delta^{13}C_{org}$ negative excursions is vary in different regions during the CPE, e.g. four negative excursions is found in western Tethys, while only one in South China (Sun et al., 2016; Dal Corso et al., 2018; Shi et al., 2019). The number of negative excursions for $\delta^{13}C_{org}$ and $\delta^{13}C_{carb}$ may also be different in the same region (Sun et al., 2016), and even instances of decoupling between the two trends (Li et al., 2020). The mechanisms causing the variation in organic carbon isotope records across different regions are also unclear. The eruption of the Wrangellia large igneous province, causing massive $CO_2$ injection into the atmosphere or oceans, may have been one of the primary reasons for these repeated negative excursions (Furin et al., 2006; Dal Corso et al., 2020; Lu et al., 2021; Mazaheri-Johari et al., 2021). However, in addition to volcanic emissions of light carbon, the substantial input of terrestrial organic matter and later diagenesis could also significantly affect marine organic carbon isotope records (Jiang et al., 2012; Li et al., 2020). Determining how many modes of carbon isotope disturbance occur during the Carnian Pluvial Episode and understanding the driving mechanisms of each is crucial for a deeper understanding of the disruptions in the carbon cycle during the Carnian.

In this study, we use newly acquired carbonate and organic carbon isotope data, as well as sedimentological data from the Carnian of the Nanpanjiang Basin and Lijiang Basin, combined with previously reported carbon isotope data, to systematically analyze changing patterns of carbon isotope curves during the Carnian. We explore the reasons for the decoupling of carbonate and organic carbon isotope data curves and assess the impact of terrestrial input on marine carbon isotope records. Current results show that influenced by terrestrial material input and carbon release from volcanism, the carbon isotope records from shallow to semi-deep marine shelf to slope settings exhibit 2 different patterns, further revealing the link between Carnian Pluvial Episode and large-scale changes in marine carbon isotope records.

## 2. Geological setting

During the Middle to Late Triassic, South China was located on the eastern margin of the Tethys Ocean, around latitude 15-30° N (Fig. 1A). The Middle-Upper Triassic marine stratigraphy of South China is mainly distributed in the Nanpanjiang Basin and the Lijiang Basin (Ma et al., 2009). From the Late Ladinian to the Carnian, the Indosinian movement caused most of the Upper Yangtze Block to uplift, while the Nanpanjiang Basin became a typical remnant marine basin in this region (Wu, 2003; Mei, 2010). The Lijiang Basin is situated to the west of the Kangdian ancient land (Fig. 1B), which is near the Sibumasu block (Chen et al., 2020).

The Xiashulao section is situated 75 km northeast of Lijiang City, Yunnan Province (GPS: N: 27°15' 59"; E:100°53'58", Fig. 1B), located at the eastern of the Lijiang Basin during the Middle-Late Triassic. It predominantly consists of the Middle Triassic Beiya Formation, Upper Triassic Zhongwo Formation, and Songgui Formation, with a total section thickness of 71 m (Fig. 2A). The top exposures of dolomitic limestone and argillaceous limestone in the Beiya Formation represent platform to open marine environments (Yin et al., 2006; Dong et al., 2013). The lower part of the Zhongwo Formation consists of thick-bedded, massive-micritic limestone, and upward the limestone beds become thinner and more muddy, reflecting the evolution of the platform into deep-water slopes (Dong et al., 2022). The top of the Zhongwo Formation is disconform with the Songgui Formation, and the bottom of the Songgui Formation is a series of yellowish-green sandy limestone, which indicating land-sea interaction (He, 1999). In the Xiashulao section, two conodont zones were identified: the *Q. polygnathiformis* Zone and the *Q. polygnathiformis - noah* assemblage Zone, marking the Julian and Tuvalian of the Carnian (Fig. 2A, see Appendix A for details). Additionally, about 1.8 m below the first appearance of *Q. polygnathiformis* in the Xiashulao section, ammonite *Xenoprotrachyceras* sp. was found (Fig. 2A, see Appendix A for details). The occurrence of *Xenoprotrachyceras* within the Ladinian ammonite *Haoceras xingyiensis* Zone, which is close to the first occurrence of *Q. polygnathiformis,* is also reported from Nimaigu section of Xingyi, where is close to Xiashulao section (Wang et al., 1998; Zou et al., 2015). This further provides evidence that the strata below the *Q. polygnathiformis* Zone are from the Ladinian.

The Erguan section is located in Pingba, which is 40 km to the northeast of Anshun (GPS: N: 27°15'59"; E:100°53'58", Fig. 1B). The section lies on the northern edge of the Nanpanjiang Basin and in a shelf-slope depositional environment (Ma et al., 2009). The Erguan section, from bottom to top, consists of the Gaicha Formation and Sanqiao Formation, which are in conformable contact, with a total thickness of 83.15 m (GPS: N: 26°24'7"; E:106°17'0", Fig. 1B). The upper part of the Gaicha Formation is comprised of dolomitic limestone and limestone, which representing a platform facies (Enos et al., 1998; Mei, 2010). The Sanqiao Formation sequence shows three cycles of transition from thin interbedded calcareous shale and limestone to medium-thick limestone, representing a shelf to deep-water slope environment (Lehrmann et al., 2005). At the Erguan section, two conodont zones were identified: *Quadralella polygnathiformis* Zone and *Q. polygnathiformis - noah* assemblage Zone, indicating the Julian and Tuvalian substages of the Carnian, respectively (Fig. 2B, see Appendix A for details). In previous study, the bottom boundary of the *Q. polygnathiformis - noah* assemblage Zone is determined with the aid of bivalve *Halobia* cf. *zitteli*, as *Q. noah* is extended into the *Quadralella polygnathiformis* Zone in some studies (Sun et al., 2016). Since no age-diagnostic Julian bivalve was found at Xiashulao and Erguan, the Julian - Tuvalian boundaries of these two sections are still unclear.

The Duanqiao section is located in Duanqiao Town of Guanling, 50 km southwest of Anshun (GPS: N: 25°53'49"; E:105°38'31", Fig. 1B). This section is settled at the northwestern Nanpanjiang Basin, and exposes the Zhuganpo Formation and the Xiaowa Formation (Fig. 2C), with a total thickness of 50.70 m. The Zhuganpo Formation consists of a set of gray nodular argillaceous limestone and argillaceous limestone, while the Xiaowa Formation features black shale and interbedded mudstone and gray-black thin-layered argillaceous banded limestone. The lithostratigraphy at Duanqiao is overall similarity with the Longchang section (Sun et al., 2016), where one conodont Zone, *Q. polygnathiformis*, identifies the Julian substage of the Carnian (Fig. 2C, see Appendix A for details).

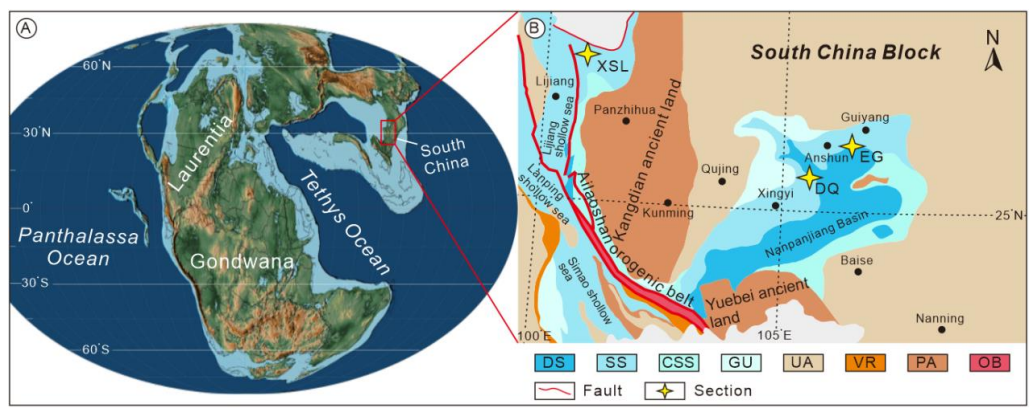

**Fig. 1.** (A) Paleogeographic map of Carnian (Late Triassic), modified from (Scotese, 2014). XSL: Xiashulao section; EG: Erguan section; DQ: Duanqiao section. (B) Carnian palaeogeographical map of South China and location of studied sections (modified from Ma et al., 2009). DS: deep sea; SS: shallow sea; CSS: coast to shallow sea; GU: gulf; UA: uplift area; VR: volcanic region; PA: paleoland; OB: orogenic belt.

## 3. Materials and Methods

In this study, sampling was carried out in the Xiashulao, Erguan, and Duanqiao sections for microfacies, conodont biostratigraphy, carbon isotope geochemistry, total organic carbon and nitrogen content, and major and trace elements analysis. A total of 126 carbonate samples and 60 conodont samples were collected from the Xiashulao section, 175 carbonate and 40 shale samples, and 97 conodont samples from the Erguan section. In the Duanqiao section, 46 carbonate samples were collected for microfacies, conodont biostratigraphy and geochemistry analysis. Due to the different thicknesses and shale proportions of Xiashulao, Erguan and Duanqiao, samples for geochemical analysis in those section were sampled at intervals of 80 cm, 50 cm and 100 cm, respectively. The conodont samples for Xiashulao, Erguan and Duanqiao were collected at intervals of 200 cm, 80 cm and 100 cm, respectively.

*3.1 Microfacies analysis and conodont sample pre-processing*

The microfacies analysis is based on the slides. All carbonate samples are cut in a direction perpendicular to the bedding plane and made into slides with area of 2×2 cm. All slides are observed and photographed under the polarizing microscope. The conodont samples were broken into about

5 cm fragments and dissolved in 6%–10% acetic acid. The acid solution was buffered with tricalcium phosphate and replaced every 48 hours until the samples were completely dissolved. The residues were washed, sieved with water, dried at 50°C, and picked under a stereoscopic microscope. The results of the microfacies analysis and the biostratigraphy are detailed in Appendix A.

*3.2 $\delta^{13}C_{carb}$-VPDB, $\delta^{18}O_{carb}$-VPDB, TOC and C/N ratios analysis*

A total of 316 carbonate samples were processed for carbon and oxygen isotope analysis. The analysis was performed using a stable isotope mass spectrometer (Thermo Fisher Scientific, GasBench II-MAT 253), following the method proposed by Song et al. (2013). For organic carbon isotope and total organic carbon (TOC) content analysis, a total of 183 samples were tested by using Delta V advantage (Thermo Fisher). In addition, 72 samples were tested in this study total nitrogen content and this analysis was conducted on an Vario MACRO cube elemental analyzer. Pretreatment and testing process of organic carbon isotope, and total carbon and nitrogen content is according to the method of Du et al. (2021). All experiments and tests were conducted at the State Key Laboratory of Biogeology and Environmental Geology, China University of Geosciences, Wuhan. The isotope ratios (δ) are reported relative to the Vienna Pee Dee Belemnite (VPDB) standard and are expressed in per mil (‰). High-purity $CO_2$ (99.999%) cylinder gas was used as the reference standard during the testing process. The detailed testing process is shown in Appendix A.

*3.3 Major and trace elements analysis*

Major and trace element analyses of 123 samples were conducted in the laboratory of Wuhan SampleSolution Analytical Technology Co. Ltd and detailed testing process is shown in Appendix A. The major element tests were performed using an Zsx Primus II wavelength dispersive X-ray fluorescence spectrometer (XRF). The sample pretreatment of whole rock major element analysis was made by melting method. The standard curve was based on the Chinese National Standard Material GBW07101-14 (National Standard of P.R. China). The data were corrected using the theoretical α coefficient method. The relative standard deviation (RSD) yielded a value of less than 2%. Trace element analysis of whole rocks was conducted on an Agilent 7700e ICP-MS. The overall testing and analysis were performed at 22.4°C and 21% humidity, according to GB/T14506.30-2010 (National Standard of P.R. China, 2011). The analytical precision of the determinations of the trace elements was at or above 2% in all cases. The concentration of Mn is calculated by taking the weight fraction of Mn and multiplying it by a coefficient (molecular weight [Mn]/molecular weight [MnO]).

## 4. Results

*4.1 $\delta^{13}C_{carb}$ and $\delta^{18}O_{carb}$*

Carbon and oxygen isotope data from the Xiashulao, Erguan, and Duanqiao sections are shown in Fig. 2. In the Xiashulao section, the $\delta^{13}C_{carb}$ values range from -0.38 to +3.06‰, with an average of +2.08‰ (Fig. 2A). The $\delta^{18}O_{carb}$ values range from -12.82 to -2.80‰, with an average of -5.48‰ (Fig. 2A). Four major negative carbon isotope excursion intervals (CIEs) are identified. At the base of the section, $\delta^{13}C_{carb}$ values are stable at about 2‰ (Fig. 2A). Starting from 15 m in the section, the $\delta^{13}C_{carb}$ curve begins to fluctuate (carbon isotope excursion near Ladinian-Carnian boundary, LCB CIE), with a negative peak value of -0.4‰ and a significant negative shift of about 2‰. Within this interval, both $\delta^{13}C_{carb}$ and $\delta^{18}O_{carb}$ values are reduced compared to other parts of the section (Fig. 2A). Two continuous negative $\delta^{13}C_{carb}$ excursion intervals, namely CIE 1 and CIE 2, occur from the upper part of the conodont *Q. polygnathiformis* Zone to the base of the *Q. polygnathiformis - noah* assemblage Zone in the Zhongwo Formation. Both peaks are around +1.2‰, with a total negative shift of about 1.5%, accompanied by a $\delta^{18}O_{carb}$ negative shift (Fig. 2A). The fourth major $\delta^{13}C_{carb}$ negative shift interval (CIE 3-4) is near 93 m in the section, showing three negative peaks with the smallest shift of about 1.2‰ and the largest shift of about 1.8‰. The trend of $\delta^{18}O_{carb}$ curve is similar (Fig. 2A).

The curve of $\delta^{13}C_{carb}$ shows that the values range from -1.51 to +2.19‰, with an average of +1.13‰ in the Erguan section (Fig. 2B). However, the $\delta^{18}O_{carb}$ values are relatively unstable, showing multiple fluctuations ranging from -10.16 to -2.38‰, with an average of -5.75‰ (Fig. 3D). The background value of $\delta^{13}C_{carb}$ is around 1.7‰, and the curve records four major CIEs. The first negative carbon isotope excursion interval (LCB CIE) starts near the first occurrence of *Q. polygnathiformis*, with $\delta^{13}C_{carb}$ peaking at 13.7 m in the section (+0.05‰) and a negative shift of around 1.6‰. From about 39 m to 45 m in the section is the second $\delta^{13}C_{carb}$ negative interval (CIE 1-2), beginning in the limestone below the black shale interval of the Sanqiao Formation with a shift of about 3.1‰. CIE 3 starts at 57 m in the section, with a negative peak of around -0.9‰ and a maximum negative shift of 2.6‰. The final $\delta^{13}C_{carb}$ negative interval (CIE 4) occurs around 69 m in the section, with carbon isotope values rapidly fluctuating to around -0.3‰, then gradually recovering to +1.7‰, with a negative shift of 2%. Within four $\delta^{13}C_{carb}$ negative intervals, $\delta^{18}O_{carb}$ also shows negative trend with similar fluctuation amplitudes. However, outside the CIE intervals, there are also five negatives shifts of the $\delta^{18}O_{carb}$ curve (Fig. 2B).

In the Duanqiao section, $\delta^{13}C_{carb}$ values range from -1.51 to +2.19‰, with an average of +1.13‰ (Fig. 2C). The $\delta^{18}O_{carb}$ values range from -10.16 to -2.38‰, with an average of -5.75‰ (Fig. 3G). The $\delta^{13}C_{carb}$ value is stable at about +2.9‰ at the base, with two major negative carbon isotope shift intervals above, located near the Zhuganpo Formation-Xiaowa Formation boundary (Fig. 2C). The two negative peaks, CIE 1 and CIE 2, with a smaller peak around +2.0‰ near 38 m in the section and a larger peak around +1.2‰ at 42 m, with a maximum negative shift of 1.7‰. The $\delta^{18}O_{carb}$ values also show a strong negative shift at the larger peak, with an amplitude of about 4‰. Due to the exposure conditions of the section, the CIE 3 is incompletely recorded, with $\delta^{13}C_{carb}$ values fluctuating from about +2.7‰ down to +1.5‰, and $\delta^{18}O_{carb}$ values relatively negative, stable at -6‰ (Fig. 2C).

*4.2 $\delta^{13}C_{org}$, TOC and C/N*

As Fig. 2A shows, $\delta^{13}C_{org}$ curve fluctuate several times in the Xiashulao section, ranging from -28.06‰ to -24.44‰, with an average value of -26.27‰. Overall, the organic carbon isotope shows a gradual positive shift along the Xiashulao section. In the LCB CIE interval, there is a decoupling between $\delta^{13}C_{org}$ and $\delta^{13}C_{carb}$, with $\delta^{13}C_{org}$ values increasing from -27.38‰ to -24.82‰. After the positive shift, the $\delta^{13}C_{org}$ values fluctuate between -28.07‰ and -26.11‰. In the CIE1-4 intervals, the peaks of $\delta^{13}C_{org}$ and $\delta^{13}C_{carb}$ are also asynchronous. Within the CIE 1-2, the $\delta^{13}C_{org}$ values first shift positively from -27.21‰ to -25.39‰, then negatively to -26.86‰. Values of organic carbon isotope fluctuate frequently in the interval CIE 3-4, ranging from -26.63‰ to -24.44‰. Most samples in the Beiya Formation of the Xiashulao section have TOC contents below 0.05%. An increase in TOC content occurs in the LCB CIE interval from the top of the Beiya Formation to the bottom of the Zhongwo Formation, with TOC content fluctuating between 0.05% and 0.10% in the overlying samples. In the upper part of the section, the TOC content changes in a trend like the $\delta^{13}C_{org}$ curve (Fig. 2A). At the top of the section, the TOC content peaks at about 0.20%. The C/N ratio in this section ranges from 4.02 to 13.80, with an average of about 9.44. The trend of C/N ratio changes is similar to that of TOC content, peaking within the CIE 3-4 (Fig. 2A).

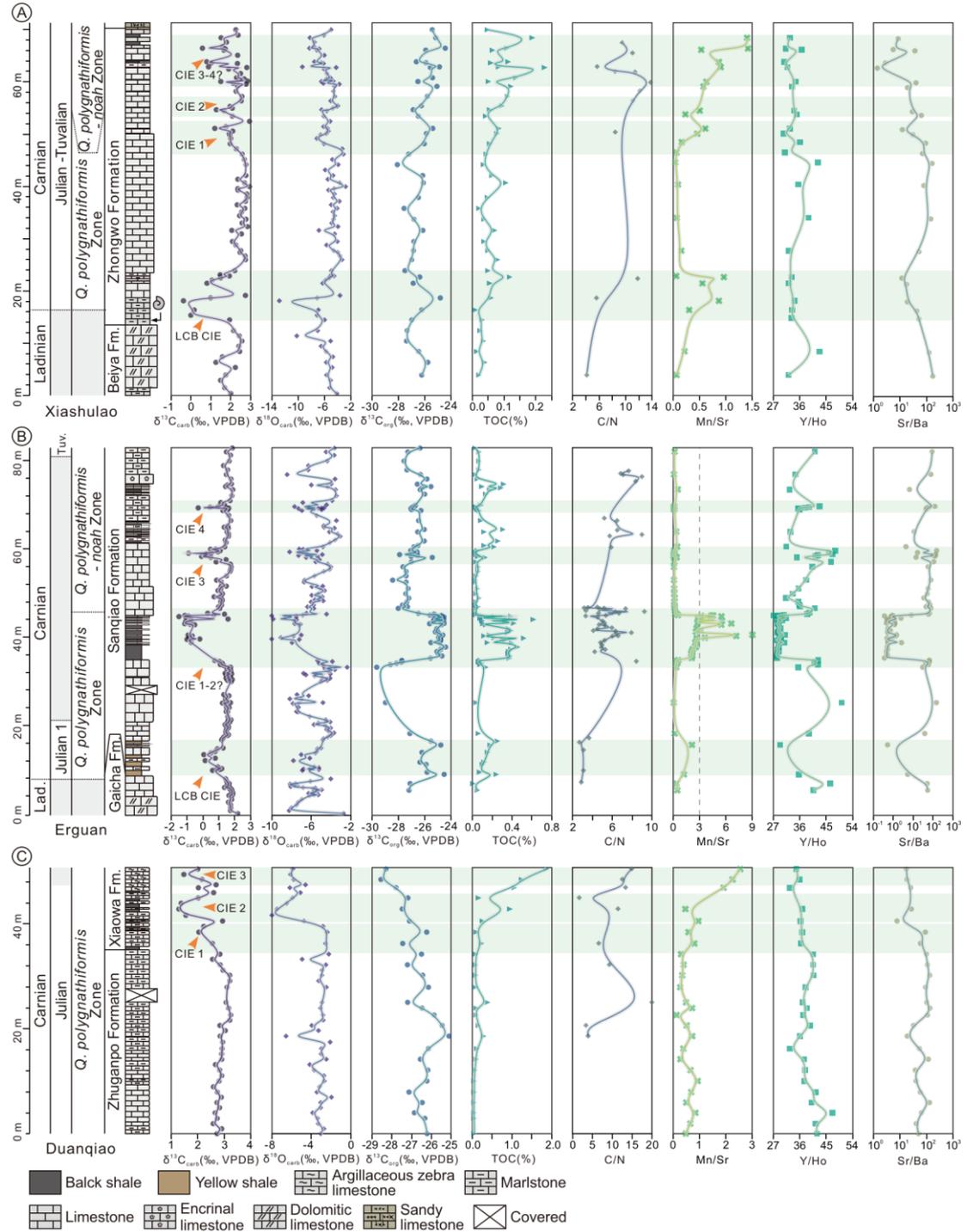

**Fig. 2.** Lithology, biostratigraphy, $\delta^{13}C_{carb}$ and $\delta^{18}O_{carb}$, $\delta^{13}C_{org}$, TOC, C/N ratios, Mn/Sr ratios, Y/Ho ratios and Sr/Ba ratios of the studied sections. (A) Xiashulao section from Lijiang Basin. (B) Erguan section from Nanpanjiang Basin. (C) Duanqiao section from Nanpanjiang Basin. Black arrows

indicate the main intervals of terrestrial inputs and enhanced water circulation. Fm.: Formation. Tuv.: Tuvalian.

In the Nanpanjiang Basin, $\delta^{13}C_{org}$ values in the Erguan section range from -29.63‰ to -24.44‰, with an average value of -26.23‰. The lower part of the curve shows two major positive excursion intervals, while the upper part shows less pronounced fluctuations (Fig. 2B). The first positive excursion interval corresponds to the LCB CIE interval, where $\delta^{13}C_{org}$ values decouple from $\delta^{13}C_{carb}$, rising from about -27.20‰ to about -24.53‰. The second positive excursion interval corresponds to CIE 1-2 interval, where $\delta^{13}C_{org}$ values also decouple from $\delta^{13}C_{carb}$, rising from -28.15‰ to -24.44‰. At the 57 m of the section (CIE 3), $\delta^{13}C_{org}$ values fluctuate between -27.9‰ and -25.5‰, generally stabilizing around -27‰ with relatively small negative shift. Around the 69 m (CIE 4), $\delta^{13}C_{org}$ values show almost no change or a very weak negative shift, fluctuating from -26.3‰ to -26.8‰. Overall, the TOC content of the limestone samples in the Erguan section is relatively low, about 0.1%. Most shale samples have TOC contents higher than 0.2%, with peaks reaching 0.6%. Samples with high TOC content are mainly from the CIE 1-2. In addition, similar to the Xiashulao section, the trend of TOC content changes in Erguan samples is consistent with the trend of the $\delta^{13}C_{org}$ curve. The C/N ratio in this interval ranges from 2.69 to 9.25, with an average of about 5.53, showing an overall increasing trend from about 3 in the lower part of the interval to about 7 in the upper part.

In the Duanqiao section, the $\delta^{13}C_{org}$ curve shows some degree of irregular fluctuation in the lower part of the section, with a maximum value of -25.1‰ and a minimum value of -27.4‰, with most data between -26‰ and -27‰. Above the 38 m of the section, $\delta^{13}C_{org}$ values dropping from -27.1‰ to -28.5‰ in the CIE intervals. Before the main carbon isotope negative excursion begins, the TOC content in limestone in the Duanqiao section is low, typically below 0.1%. During the carbon isotope negative excursion interval, the TOC content in limestone rapidly increases from about 0.1% to about 1.9%. The C/N ratio in this section ranges from 3.91 to 19.99, with an average value of approximately 9.69, with most C/N ratios fluctuating between 3 and 15. Within the $\delta^{13}C_{carb}$ negative excursion interval, the overall trend of C/N ratio change is similar to that of TOC content, showing a gradual increase.

*4.3 Major and trace element*

At Xiashulao section, Manganese values are ranging from ~30 to 660 ppm. In most of samples, the Manganese values is less than 225 ppm. The value increases twofold (350-429 ppm) at the base of the Zhongwo Formation (17-22m of the section). In addition, Manganese value also increases notably at the top of Zhongwo Formation, to the max values of 660 ppm. Mn/Sr ratios show a same trend, ranging from 0.31 to 0.96 at the base of the Zhongwo Formation, and 0.23 to 1.42 in the interval 48-68 m of the Xiashulao section (Fig. 2A). Y/Ho ratios remain ~33 in the LCB CIE interval and CIE 1-4 interval. Before the LCB CIE and in the middle part of Zhongwo Formation, the Y/Ho ratio reach to 42. The trend of Sr/Ba ratio is similar to that of Y/Ho, the Sr/Ba ratios are lower in the LCB CIE and CIE 1-4 intervals (10-50), and reach to 163 in the middle part of Zhongwo Formation (Fig. 2A).

The concentration of Mn ranges from ~80 to 1800 ppm at Erguan section, and the average value is 385 ppm. The range of variation of the Mn/Sr ratio is from 0.09 to 9. In the LCB CIE interval, Mn/Sr ratios increase from ~0.5 to 2 (Fig. 2B). Subsequently, this value declines to about 0.30 in the lower part of Sanqiao Formation. The second peak of Mn/Sr ratio is in the CIE1-2 interval, with the value of 9. At the boundary of Julian and Tuvalian, the Mn/Sr ratio decreases rapidly to about 0.57. The value then fluctuates around 0.2 in the upper part of Sanqiao Formation. Multiple fluctuations in Y/Ho values, ranging from 27 to 50. Y/Ho ratio decreased to ~29 in the LCB CIE interval and ~27 in CIE 1-2 interval. In contrast, in interval CIE 3 and CIE 4, the Y/Ho values are higher, at ~42 and ~38, respectively. The Sr/Ba ratio of the Erguan section varies in a wide range, from 0.35 to about 150. The trend is similar to that of the Y/Ho ratio, with low values (most <1) falling within the CIE1-2 interval. In the intervals CIE3 and CIE 4, the Sr/Ba ratios are relatively high, with most values greater than 50, and close to those of the rest of the section (Fig. 2B).

In the Zhuganpo Formation of Duanqiao section, the Mn/Sr ratio fluctuates around 0.5 and increases rapidly (0.5 to 2.5) at the translation of the Zhuganpo and the Xiaowa Formation (Fig. 2C). The Y/Ho ratio shows an overall trend of slowly decreasing, ranging from 47 to 32, with the mean value of ~38. At the bottom of the section, the value of Y/Ho fluctuates around 42, while at the top of the section, most of the data is between 35-37. In Zhuganpo Formation, the Sr/Ba ratios Sr/Ba values fluctuate modestly, from 17 to about 132 (Fig. 2C). The lowest value of Sr/Ba ratio (~7) is recorded in the Xiaowa Formation, accompanied by a slight decrease in the Sr/Ba ratio (mean values from ~78 to ~19).

## 5. Discussion

*5.1. Preservation of primary carbon isotopic signals*

The distribution ranges of $\delta^{13}C_{carb}$ and $\delta^{18}O_{carb}$ values for the Xiashulao, Erguan, and Duanqiao sections mostly exceed the range of Carnian brachiopod shell $\delta^{13}C_{carb}$ (-4.25 to -1.45‰) and $\delta^{18}O_{carb}$ values (-4.00 to -0.75‰) (Korte et al., 2005; Jin et al., 2018). Diagenesis and the recrystallization of sediments can alter the δ$^{13}$C composition of carbonates, causing them to deviate from the original marine signals (Jiang et al., 2012; Swart, 2015; Jin et al., 2018; Du et al., 2021; Jin et al., 2022). This study evaluates the potential impacts of these processes on the δ$^{13}$C$_{carb}$ and δ$^{13}$C$_{org}$ records of the Xiashulao, Erguan, and Duanqiao sections based on δ$^{13}$C$_{carb}$ - δ$^{18}$O$_{carb}$ correlation diagrams, δ$^{13}$C$_{org}$ - TOC correlation diagrams, and Mn/Sr ratios.

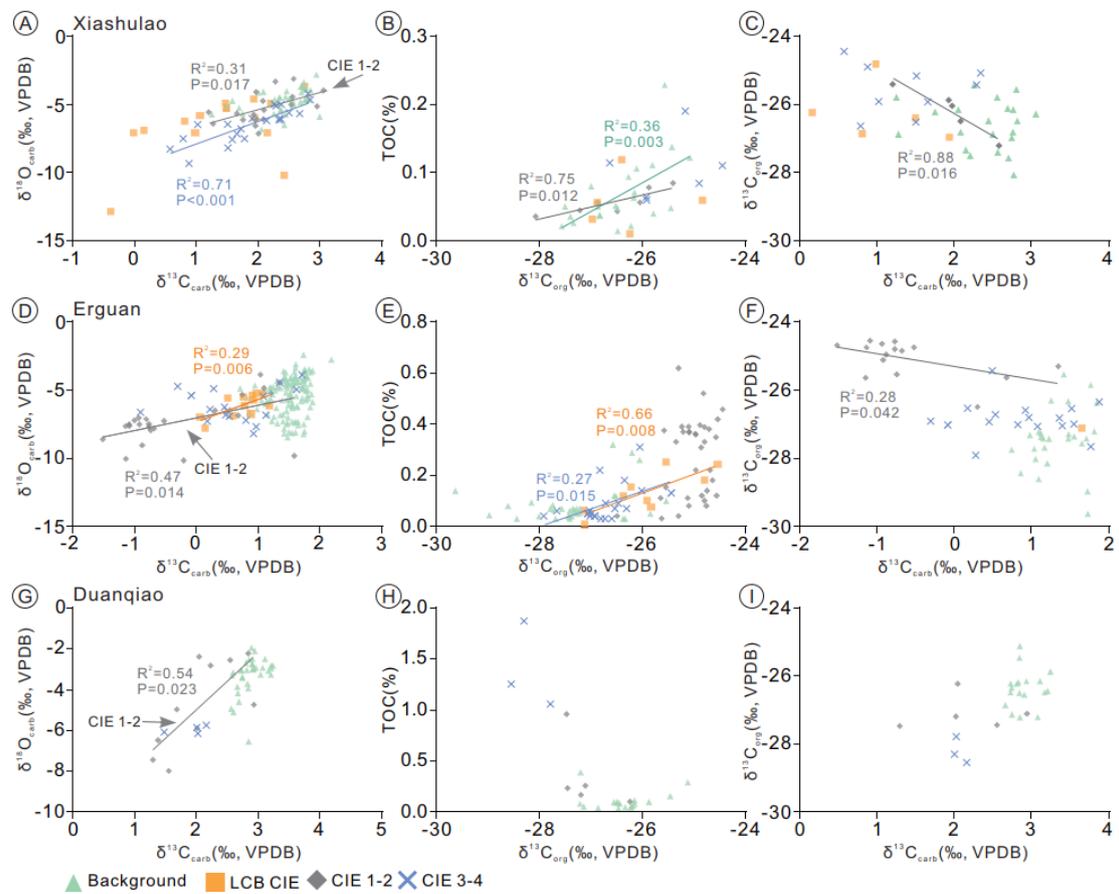

**Fig. 3.** Cross plots of δ$^{13}$C$_{carb}$ and δ$^{18}$O$_{carb}$, δ$^{13}$C$_{org}$ and TOC, and δ$^{13}$C$_{carb}$ and δ$^{13}$C$_{org}$ data from the studied sections. (A) Cross plots of δ$^{13}$C$_{carb}$ and δ$^{18}$O$_{carb}$ of Xiashulao section. (B) Cross plots of δ$^{13}$C$_{org}$ and TOC of Xiashulao section. (C) Cross plots of δ$^{13}$C$_{carb}$ and δ$^{13}$C$_{org}$ of Xiashulao section. (D) Cross plots of δ$^{13}$C$_{carb}$ and δ$^{18}$O$_{carb}$ of Erguan section. (E) Cross plots of δ$^{13}$C$_{org}$ and TOC of Erguan section. (F) Cross plots of δ$^{13}$C$_{carb}$ and δ$^{13}$C$_{org}$ of Erguan section. (G) Cross plots of δ$^{13}$C$_{carb}$

and $\delta^{18}O_{carb}$ of Duanqiao section. (H) Cross plots of $\delta^{13}C_{org}$ and TOC of Duanqiao section. (I) Cross plots of $\delta^{13}C_{carb}$ and $\delta^{13}C_{org}$ of Duanqiao section.

Previous studies have shown that the carbon and oxygen isotopic composition of sedimentary carbonates can be altered through diagenetic interactions with fluids, resulting in a positive correlation between carbon and oxygen isotopes (Knauth and Kennedy, 2009; Derry, 2010)). However, when distinguishing the background layers from the negative excursion intervals in the study sections, there is no significant correlation between $\delta^{13}C_{carb}$ and $\delta^{18}O_{carb}$ in the event background layers of the three sections. In the Xiashulao section, the $\delta^{13}C_{carb}$ and $\delta^{18}O_{carb}$ data of samples from CIE 1-4 interval exhibit significant correlations (Fig. 3A). The strong correlation of $\delta^{13}C_{carb}$ and $\delta^{18}O_{carb}$ values in the Erguan section mainly appears in the intervals LCB CIE I and CIE 1-2 (Fig. 3D), while the Duanqiao section show a not very significant correlation in CIE 1-2 interval (Fig. 3G).

The correlation between $\delta^{13}C_{carb}$ and $\delta^{18}O_{carb}$ may suggest the burial or atmospheric diagenetic origin of the negative excursion signal (Knauth and Kennedy, 2009; Derry, 2010). However, this interpretation is inconsistent with the following facts: (1) the main depositional structure of the samples in the study sections are well-preserved (Appendix A); (2) the observed patterns of $\delta^{13}C_{carb}$ variations are similar across the South China region and can be compared with $\delta^{13}C_{carb}$ curves from other global regions (Fig. 4). Additionally, the Mn/Sr ratios shows that diagenetic modification has a negligible effect on sample from Xiashulao, Erguan and Duanqiao. Mn/Sr ratio is commonly used to evaluate diagenetic alteration of carbonates (Banner, 1995; Jacobsen and Kaufman, 1999). In general, Mn/Sr ratios less than 3 indicate that a sample has not been significantly altered by diagenesis and can be used to indicate the original isotopic composition of seawater, and the threshold of 10 is widely used to distinguish weakly altered from strongly altered carbonates (Kaufman and Knoll, 1995; Knauth and Kennedy, 2009; Ke et al., 2023). In all sample from Xiashulao and Duanqiao, and most sample from Erguan section, the Mn/Sr ratios of most samples is below 3 (Fig. 2). Only nine samples in the CIE 1-2 interval of the Erguan section show Mn/Sr ratios exceeds 3 and below 10 (Fig. 2B), consistent with only limited diagenetic effect. Furthermore, the occurrence of low $\delta^{18}O_{carb}$ values in CIE 1-2 interval might be associated with enhanced freshwater inflow into the sediments during CPE, because $\delta^{18}O_{carb}$ is sensitive to water-rock interactions (Banner and Hanson, 1990). Considering all the above, the positive correlation between $\delta^{13}C_{carb}$ and $\delta^{18}O_{carb}$ may not be a result of diagenesis (Husson et al., 2015).

The composition of $\delta^{13}C_{org}$ might be altered during diagenesis due to anaerobic oxidation or thermal maturation processes (Jiang et al., 2010). Given the low TOC content in carbonate rocks, their $\delta^{13}C_{org}$ values are more likely to be influenced by thermal maturation (Jiang et al., 2012). Thermal maturation decreases TOC content and leads to a negative correlation between $\delta^{13}C_{org}$ and TOC values (Hayes et al., 1989; Clayton, 1991; Tocqué et al., 2005). In the Xiaoshulao and Erguan sections, $\delta^{13}C_{org}$ values shift positively with increasing TOC content (Fig. 3B, E), while in the Duanqiao section, there is no significant correlation between TOC content and $\delta^{13}C_{org}$ values (Fig. 3H), indicating that $\delta^{13}C_{org}$ values are not significantly affected by thermal maturation. It is noteworthy that samples with high TOC content mainly originate from the CPE event strata, and previous studies have shown a significant increase in TOC content in sediments post-event (Zhang et al., 2022; Lukeneder et al., 2024). Thus, the correlation between $\delta^{13}C_{org}$ values and TOC content in the three sections (Fig. 3B, E, H) is more likely a reflection of the CPE rather than diagenetic alteration.

*5.2. Stratigraphic correlation of studied sections*

Multiple negative shifts of $\delta^{13}C_{carb}$ during the Carnian Pluvial Episode have been widely accepted as markers for stratigraphic correlation in South China (Sun et al., 2016; Zhang et al., 2017; Rigo et al., 2018; Li et al., 2020). In this study, in the Xiashulao and Erguan sections, intervals near the first appearance of the conodont *Quadralella polygnathiformis* show the simultaneous occurrence of negative $\delta^{13}C_{carb}$ excursions and positive $\delta^{13}C_{org}$ values (Fig. 2A, B). This trend has not been previously reported in studies of $\delta^{13}C_{carb}$ and conodont biostratigraphy in South China (Sun et al., 2016; Sun et al., 2019; Zhang et al., 2022). Some studies have suggested a negative $\delta^{13}C_{carb}$ isotopic shift near the Ladinian-Carnian boundary (Zhang et al., 2021; Lestari et al., 2024). Korte et al. (2005) suggest that trends of late Ladinian $\delta^{13}C_{carb}$ curve turn to negative in whole rock carbon isotopes. Tests of $\delta^{13}C_{carb}$ in limestones and marls at the top of the Ladinian stage in the Monte San Giorgio section in Switzerland also show a negative excursion of about 2-4‰ (Stockar et al., 2013). In addition, a negative excursion of about 2‰ near the Ladinian-Carnian boundary has also been found in the Augusta Mountain Formation in Nevada, North America (Bonuso et al., 2018). Previous studies suggest that the negative excursion may be showed near the first appearance of the conodont *Quadralella polygnathiformis*. Thus, the first occurrence of *Quadralella polygnathiformis* showed above the LCB CIE is indicating the onset of the Carnian (Fig. 2A, B).

In the western Tethys, the Carnian Julian 2 records two major negative carbon isotope excursion intervals, with one distinct negative excursion occurring in both Tuvalian 1 and Tuvalian 2 (Dal Corso et al., 2018). However, in the eastern Tethys, Julian 2-Tuvalian 2 often corresponds to a very thin stratigraphic thickness due to slower sedimentation rates (Dal Corso et al., 2024).

Because of the sampling bias and other issues, the two carbon isotope negative excursions of Julian 2 may be recorded as a big one or two continuous but indistinguishable peaks in δ¹³C$_{carb}$ curves (Zhang et al., 2022).

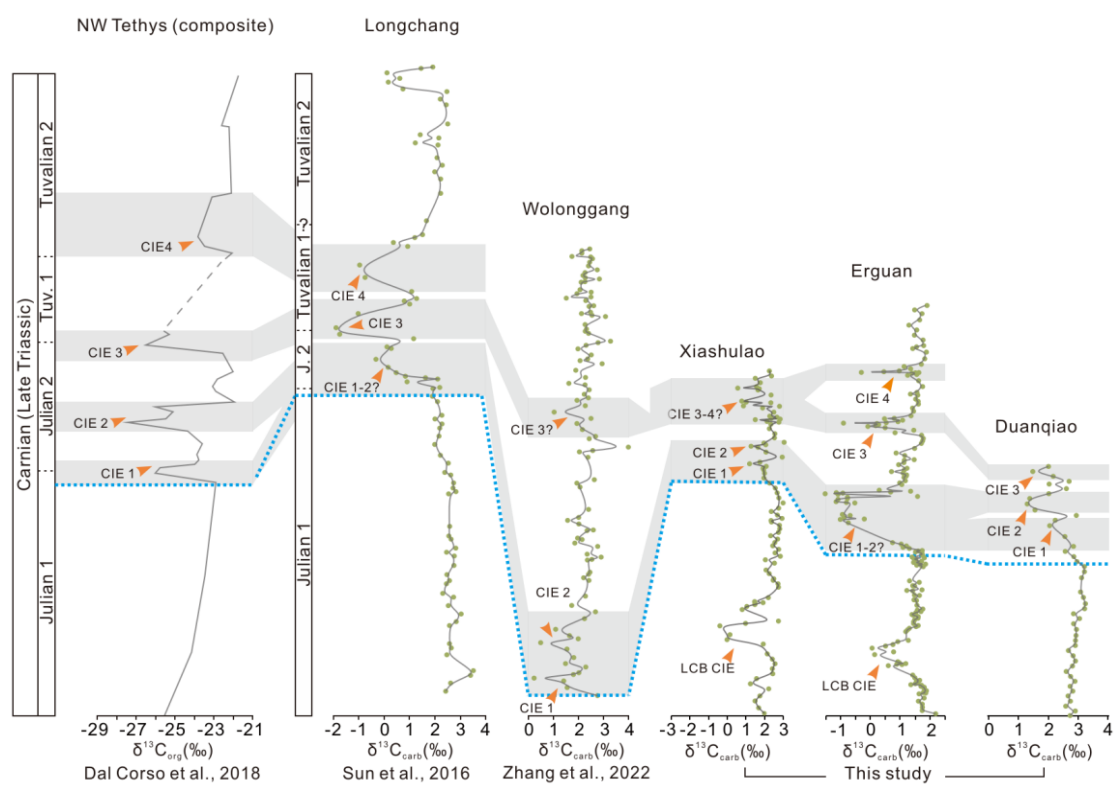

**Fig. 4.** Carbonate carbon isotope chemostratigraphic correlation across the CPE in western Tethys and eastern Tethys (data from this study and Dal Corso et al., 2018; Zhang et al., 2022).

In this study, Xiashulao shows one δ¹³C$_{carb}$ negative excursion interval with two continuous peaks, similar to the Wolonggang section (Fig. 4). In the Erguan section, the first two δ¹³C$_{carb}$ negative excursions of Julian 2 are difficult to distinguish, similar to records in Oman and Longchang (Sun et al., 2016; Sun et al., 2019). The third negative shift of Carbon isotope during CPE is often used as an assistant marker to distinguish the boundary of Julian and Tuvalian (Dal Corso et al., 2018; Lu et al., 2021). However, frequent carbon isotope fluctuations within the CIE 3-4 intervals at the Xiashulao section make it difficult to divide Tuvalian 1 and Julian 2 (Fig. 4). In the Erguan section, two distinct δ¹³C$_{carb}$ negative excursions of Tuvalian are recorded, which can be compared with the Longchang and western Tethys Ocean records (Sun et al., 2016; Dal Corso et al., 2018), and Tuvalian 1 and Tuvalian 2 can be distinguished by the CIE 3 and CIE 4 (Fig. 4). The Duanqiao section is similar to the previously reported Longchang section (Sun et al., 2016), where the stable interval of $\delta^{13}C_{carb}$ of Julian 1 and the two negative excursion intervals of Julian 2 and one of Tuvalian 1 can be observed (Fig. 4).

*5.3. Sources of sedimentary organic matter*

Negative $\delta^{13}C_{org}$ excursions in the Carnian have been interpreted to reflect phased input of isotopically light carbon from the Wrangellia LIP and other contemporaneous volcanism (Sun et al., 2016; Dal Corso et al., 2018; 2020; Lu et al., 2021; Mazaheri-Johari et al., 2021). However, different to previous studies, the $\delta^{13}C_{org}$ curve of the Xiashulao section shows a positive trend during CIE 1-4 interval of CPE. In the Xiashulao and Erguan, the $\delta^{13}C_{org}$ generally remains stable at around -27‰ (Fig. 2), while positive excursions in $\delta^{13}C_{org}$ are observed during the CPE, with maximum values reaching about -24.5‰. In contrast, $\delta^{13}C_{org}$ show negative shift during CPE at the Duanqiao section, which is same as previous studies (Sun et al., 2016; Dal Corso et al., 2018).

In addition to volcanism, the component of $\delta^{13}C_{org}$ might also be affected by different sources of organic matter, two of the main factors are detrital organic carbon and DOC (Jiang et al., 2012; Gu et al., 2024). It has been shown that the influx of terrestrial plant organic matter could cause a positive shift in organic carbon isotopes to -25‰ in Middle Triassic (Forte et al., 2022). Inputs of terrestrial detrital organic carbon may also have been important in influencing the isotopic composition during the CPE, for widespread of fossilized driftwood in the Carnian marine strata (Wang et al., 2008; Hamad et al., 2014; Dal Corso et al., 2015; Dal Corso et al., 2018). In addition, biomarker analyses of sediments from the Julian substage at Polzberg, Austria, indicate a significant increase of terrestrial plant components in TOC during Julian 2 (Lukeneder et al., 2024). It is less likely that DOC consolidation will result in a positive organic carbon shift at Xiashulao and Erguan than the effect of terrestrial inputs. This is because $\Delta^{13}C_{(carb-org)}$ values are in the range of 23-32‰, represents a typical photosynthetic organic matter, instead of reworked by methanotrophic bacteria (Gao et al., 2020) In order to evaluate whether there is a relationship between the positive shifts in $\delta^{13}C_{org}$ and the terrestrial inputs, the C/N ratio and Y/Ho ratio of the samples were tested in this study.

The C/N ratio, which is influenced by both terrestrial and marine organic matter, serves as an indicator of the sources of organic matter (Meyers, 1997; Sampei and Matsumoto, 2001). It is typical for deep-sea sediments to have a high inorganic nitrogen content, which results in C/N ratios that are typically less than 4 (Müller, 1977). The C/N ratios of debris from terrestrial higher plants are typically greater than 15 (Hedges et al., 1988; Orem et al., 1991). Similarly, an increase in C/N values may be indicative of alterations in the sources of organic matter. In both the Xiashulao and Erguan, there is a discernible upward trend in C/N values (from ~3 to ~10). In contrast, samples from Duanqiao section does not exhibit a notable increase in C/N values, which may suggest the influence of terrestrial organic debris on the samples from the Xiashulao and Erguan. Although the C/N ratios do not reach those typical of terrestrial sediments, the elevated trend is indicative of the presence of terrestrial inputs. In addition, the Y/Ho ratio are also employed to recognize seawater and the influx of freshwater or terrestrial inputs (Johannesson and Burdige, 2007; Siahi et al., 2018; Huang et al., 2022). Y/Ho ratios in seawater are often high (>44), but low (~28) in terrigenous materials and volcanic ash (Taylor and McLennan, 1985; Ke et al., 2023; Sreenivasan et al., 2024).

The mean value of Y/Ho ratios for Xiashulao section remain low (~33) in CIE 1-4 interval, especially in the CIE I interval, it reached a minimum value of 30 (Fig. 2A). This corresponds to an overall positive trend in $\delta^{13}C_{org}$. At Erguan, Y/Ho ratio decreased to ~27 in CIE 1-2 interval, and recovery to 38-42 in CIE 3-4 interval (Fig. 2B). This corresponds to a clear $\delta^{13}C_{org}$ positive shift in the CIE 1-2 interval and very week negative shifts in the CIE 3-4 interval (Fig. 2B). In addition, the decreasing in Y/Ho ratio also occurred together with elevated TOC content at Xiashulao and Erguan (Fig. 2A, B). This may suggest that the terrestrial source of these TOC properties. The Y/Ho ratio decreases only slightly in the CPE interval of the Duanqiao section, from 39 to about 36, which suggests that sediments may have been affected by a relatively small amount of terrestrial inputs. This is consistent with the absence of positive shift in $\delta^{13}C_{org}$ of Duanqiao section. Although an increase in Toc content was observed in the Duanqiao section during the CPE. However, changes in Y/Ho suggest that this increase may not be due to terrestrial inputs. It has been suggested that anoxia occurred in the interior of the Nanpanjiang Basin during the CPE, which the rapid increase in TOC content at Duanqiao section may be related to (Sun et al., 2016; Zhang et al., 2022).

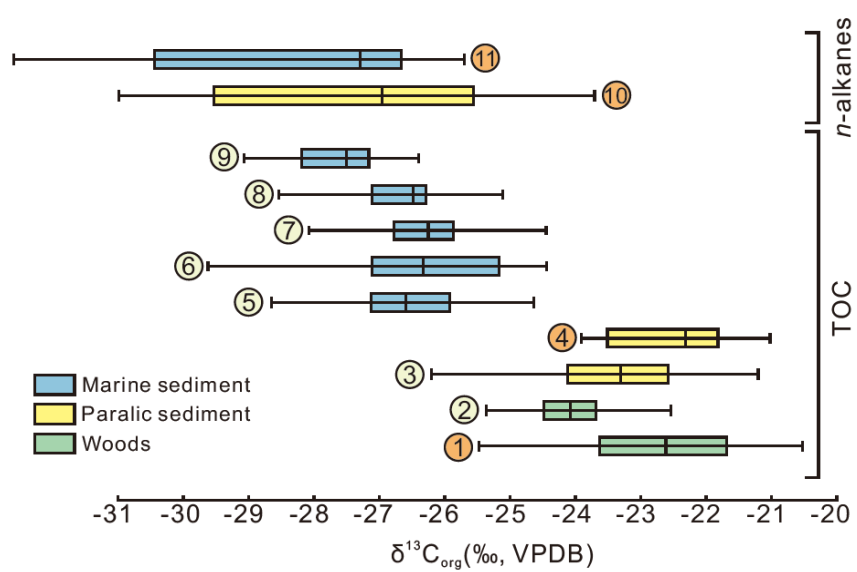

**Fig. 5.** Comparison of $\delta^{13}C_{org}$ values of the Carnian bulk-rock, woods and n-alkanes from western

and eastern Tethys. 1: Dolomites (Dal Corso et al., 2012); 2: Sichuan Basin (Jin et al., 2018); 3: Zigui Basin (Li et al., 2022); 4. Dolomites (Dal Corso et al., 2018); 5. Sichuan Basin (Jin et al., 2018); 6. Erguan, this study; 7. Xiashulao, this study; 8. Duanqiao, this study; 9. Longchang (Sun et al., 2016); 10. $^{13}C_{n-25-31}$, (Dal Corso et al., 2012); $^{13}C_{n-17-19}$, (Dal Corso et al., 2012). The boxes represent 25 and 75% quartiles, the whiskers depict 10 and 90% ranges, and the central lines in the box depict the medians.

In the Xiashulao and Erguan, the positive excursions in $\delta^{13}C_{org}$ are observed with maximum values reaching about -24.5‰, which is similar to the $\delta^{13}C_{org}$ values of samples from nearshore environments or wood fossils (Fig. 5). The values of $\delta^{13}C_{org}$ source from marine would be lower, about 27‰, close to the $\delta^{13}C_{org}$ of Duanqiao section (Fig. 5). This evidence further illustrates the relationship between positive shift of $\delta^{13}C_{org}$ and the terrestrial inputs.

*5.4 Implications for intensification of the hydrological cycle during the CPE*

CPE is characterized by the enhancement of the hydrological cycle, which may cause the increase in the terrestrial siliciclastic input (Simms and Ruffell, 1989; Arche and Lopez-Gomez, 2014). In this study, evidence for positive organic carbon isotope excursions isochronous with increased terrestrial inputs and enhanced hydrological cycle has also been found.

Paleosalinity is an important marker used to interpret the freshwater supply in marine systems (Abanda and Hannigan, 2006; Remírez et al., 2024). The Sr/Ba ratio is often used as an indicator of the variation of paleosalinity in water column (Deng and Qian, 1993; López-Buendıa et al., 1999; Cao et al., 2023), because $Ba^{2+}$ is more readily precipitated in the marine systems compared to $Sr^{2+}$ (Davis et al., 2003; Tribovillard et al., 2006; Krabbenhöft et al., 2010). A high Sr/Ba ratio is indicative of a high water column salinity. In general, Sr/Ba ratios greater than 1.0 are characteristic of seawater, while ratios below 0.5 are typical of freshwater (Deng and Qian, 1993). However, research has also proposed a limit between freshwater and marine conditions of 0.2 and 0.5 (Remírez and Algeo, 2020; Wei and Algeo, 2020). Most of the samples from the Xiashulao, Erguan and Duanqiao have Sr/Ba ratios greater than 1, revealing that these samples formed in the marine environments. However, the Sr/Ba values of samples from Xiashulao and Erguan decreases considerably during CPE, from 163 to 1, and 57 to 0.35, respectively (Fig. 2A, B). Previous studies show that the Sr/Ba ratio usually does not change by more than one order of magnitude before the CPE (Ke et al., 2023). Thus, the changes in Sr/Ba ratios during CPE indicate the enhancement of freshwater influx.

Sr/Ba ratios and Y/Ho ratios show similar trends at Xiashulao and Erguan, demonstrating the link between increased terrestrial inputs and hydrological cycle (Fig. 2A, B). In contrast to Xianshulao and Erguan, the Sr/Ba and Y/Ho ratios of samples from Duanqiao section are not significantly changed (Fig. 2C). This may indicate that the section was less impacted by terrestrial inputs and fresh water during the CPE. During CPE, Multiple enhancements of the hydrological cycle which are isochronous with the carbon isotope shifts and terrestrial inputs was reported (Ke et al., 2023). However, the peaks of Sr/Ba only observed in CIE intervals 1 and 3-4 at Xiashulao section (Fig. 2A). At Erguan section, the most significant terrestrial inputs, carbon isotope shifts and enhancement of freshwater influx occur mainly in the interval CIE1-2 (Fig. 2B).

*5.5. Decoupling of the $\delta^{13}C_{carb}$ and $\delta^{13}C_{org}$ controlled by sedimentary environment*

Two primary patterns of variation are evident in the $\delta^{13}C_{carb}$ and $\delta^{13}C_{org}$ curves during CPE

events, coupling or decoupling (Fig. 6A). In this study, the $\delta^{13}C_{carb}$ and $\delta^{13}C_{org}$ values of the Xiashulao and Erguan sections are generally negatively correlated (Fig. 3C, F). At the end of early Carnian (Julian 2), the $\delta^{13}C_{carb}$ showed a negative excursion of 2-3‰, while the $\delta^{13}C_{org}$ exhibited a positive excursion of about 3-4‰. During CPE, $\delta^{13}C_{carb}$ is considered a more stable indicator for inter-regional correlation compared to $\delta^{13}C_{org}$, which mainly controlled by the changes in DIC (Li et al., 2020; Zhang et al., 2022). A shift in the composition of organic matter caused by enhanced terrestrial inputs could greatly influence trends of $\delta^{13}C_{org}$, resulting in regional decoupling of $^{13}C_{carb}$ and $\delta^{13}C_{org}$ curves (This study and Li et al., 2020).

At Duanqiao, the $\delta^{13}C_{carb}$ and $\delta^{13}C_{org}$ curves exhibit negative excursions (Fig. 3I). However, there is a phenomenon of difficulty in fully corresponding to the number of negative excursions (Fig. 6B). The comprehensive $\delta^{13}C_{org}$ curve in the western Tethys Ocean records four distinct negative excursions: two in Julian 2 and one each in Tuvalian 1 and 2 (Dal Corso et al., 2020). In contrast, the $\delta^{13}C_{carb}$ curve of the Aghia Marina section in the same region records only three negative excursions (Muttoni et al., 2014). In South China, the Longchang section's $\delta^{13}C_{carb}$ curve records four negative excursions, whereas the main negative excursion of $\delta^{13}C_{org}$ is only one (Sun et al., 2016). This is similar to the records of the Duanqiao section in this study.

The records of $\delta^{13}C_{carb}$ and $\delta^{13}C_{org}$ decoupling are primarily sourced from the eastern Tethys Ocean (Fig. 6C), with the majority originating from the Xiashulao section, the Erguan section in this study, and the Qingyangou section (Li et al., 2020). The precise timing of the decoupling process remains unclear. In the Erguan section, decoupling is observed primarily within the Julian 2 negative excursion interval of the CPE. In the Qingyangou section, decoupling occurs mainly during the Tuvalian (Li et al., 2020). In the Xiashulao section, an overall negative correlation trend is evident throughout the CPE event.

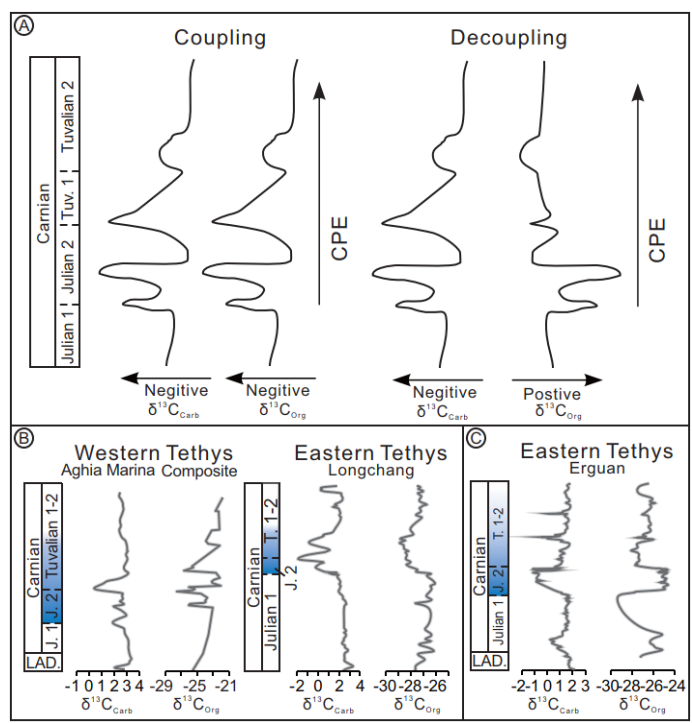

**Fig. 6.** Patterns of CPE $\delta^{13}C_{carb}$ and $\delta^{13}C_{org}$ curves. (A) Illustration of the main $\delta^{13}C_{carb}$ and $\delta^{13}C_{org}$ curves patterns. (B) Representatives of coupling pattern: western Tethys, $\delta^{13}C_{carb}$ curve from Aghia Marina (Muttoni et al., 2014) and the composite $\delta^{13}C_{org}$ curve of western Tethys (Dal Corso et al., 2018; eastern Tethys, Longchang (Sun et al., 2016). (C) Representative of decoupling pattern: eastern Tethys, Erguan (this study).

The sedimentary environment may be the main factor affecting the amount of terrestrial inputs. The discrepancy between the two signals has been proposed as being indicative of changes in lithology in CPE (Li et al., 2020; Zhang et al., 2022). However, the transition from carbonate to shale of mudstone occurs within the CIE1-2 intervals at both Erguan and Duanqiao, but the decoupled $\delta^{13}C_{carb}$ and $\delta^{13}C_{org}$ only found at Erguan. Within the Nanpanjiang basin, coupling of the $\delta^{13}C_{carb}$ and $\delta^{13}C_{org}$ curves was also identified in the previously studied Longchang and Wolonggang sections (Sun et al., 2016; Zhang et al., 2022). Similar to Duanqiao, Longchang and Wolonggang are situated within the interior of the Nanpanjiang Basin, with observed of filamenatous floatstone, showing densely packed thin shell bivalves with low content of terrestrial siliceous debris (Appendix. A). That suggest those sections receive a few terrigenous inputs. The Xiashulao section is situated in a shallow shelf environment, while the Erguan section is in an upper slope environment, both of which are in closer to the land in comparison to Duanqiao. Especially at the Erguan section, there is a significate increase in terrestrial siliceous in the CIE 1-2 interval (Appendix. A). Similarly, previous studies have also identified the decoupling of the $\delta^{13}C_{carb}$ and $\delta^{13}C_{org}$ curves in the Qingyangou section, which is situated in the slope of the Sichuan Basin (Jin et al., 2018; Li et al., 2020).

In terrestrial or open ocean records of other regions, significant negative $\delta^{13}C_{org}$ excursions during the CPE events are typically evident (Fig. 7A), occurring concurrently with the enrichment of the element mercury (Hg) (Lu et al., 2021; Jin et al., 2023). It is generally accepted that negative carbon isotope excursions are caused by the intermittent input of light carbon from volcanic activities, such as the Wrangelia Large Igneous Province (LIP) (Dal Corso et al., 2020; Sun, 2016, 2019; Mazaheri-Johari et al., 2021). However, in marine sediments from platforms or slopes, the information represented by $\delta^{13}C$ may exhibit characteristics of both terrestrial and marine organic matter (Sun et al., 2016; Zhang et al., 2022). Additionally, diagenesis may amplify the differences in isotopic records of marine and terrestrial organic carbon (Jiang et al., 2012; Gu et al., 2024), potentially leading to decoupling of organic and carbonate carbon isotope records.

Most current sedimentary records from upper inner shelf environments in proximity to land originate from the western Tethys Ocean (Fig. 7B), where no decoupling of $\delta^{13}C_{carb}$ and $\delta^{13}C_{org}$ curves was observed. In these regions, the fluctuation range of $\delta^{13}C_{org}$ values is -26‰ to -22‰,

which is relatively more positive compared to the eastern Tethys Ocean records and closely aligns

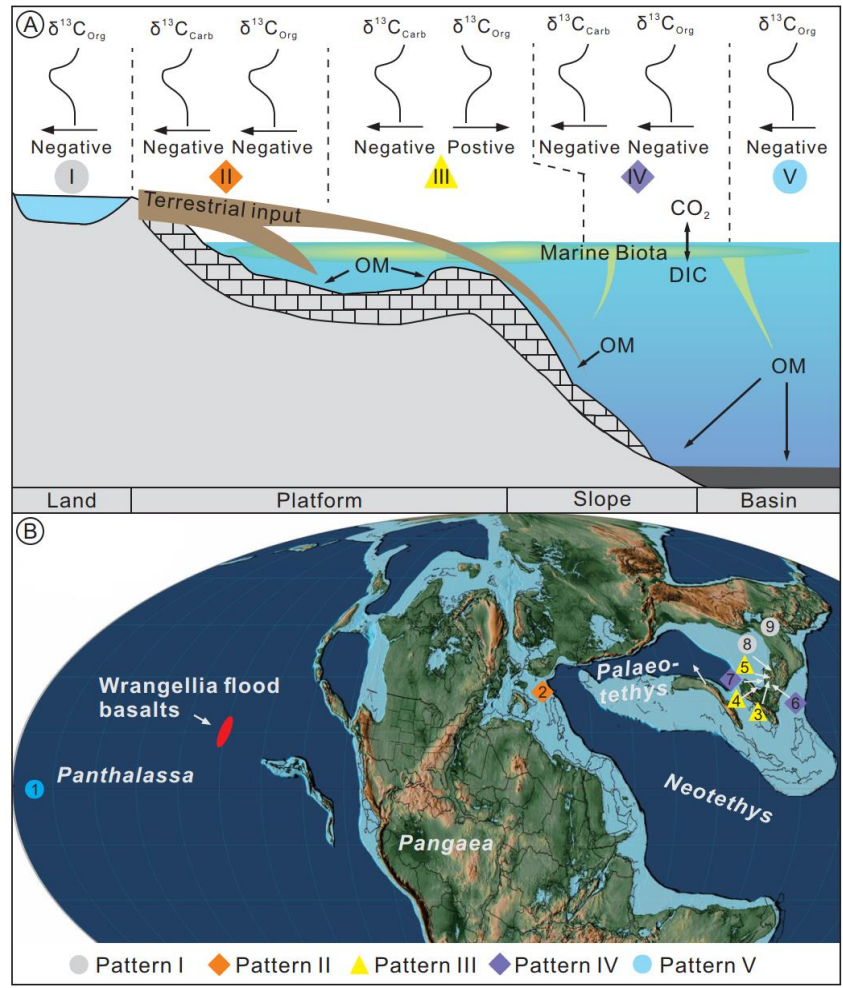

**Fig. 7.** Differences in the carbon isotope record during CPE in different paleogeographic settings. (A) Illustration of the main $\delta^{13}C$ curves patterns, pattern I: land, only negative organic carbon isotope excursion; pattern II: upper shelf, negative carbon isotope excursion; pattern III: Lower shelf to upper slope, decoupling of $\delta^{13}C_{carb}$ and $\delta^{13}C_{org}$; pattern IV: lower slope, negative carbon isotope excursion; pattern V pelagic, only negative organic carbon isotope excursion. (B) Distribution of records in different patterns with relative positions of Wrangellia LIP (Dal Corso et al., 2020): 1. Section N-O, Japan (Tomimatsu et al., 2021). 2. western Tethys (Dal Corso et al., 2018). 3, 4. Erguan and Xiashulao section, South China (this study). 5. Qingyangou (Li et al., 2020). 6. Longchang (Sun et al., 2016), and Duanqiao (this study), South China. 7. Ma' antang, South China (Shi et al., 2019).

8. Luojiagou, South China (Li et al., 2022). 9.Jiyuan, North China (Lu et al., 2021).

with the four common negative $\delta^{13}C_{org}$ excursions recorded on land (Dal Corso et al., 2020; Lu et al., 2021). Furthermore, the sedimentation rates in the western Tethys Ocean increased to 130% of their original rate during the CPE (Dal Corso et al., 2024). Additionally, multiple sections demonstrate the development of sandstone and conglomerate formations (Dal Corso et al., 2018), indicating a greater influence from terrestrial clastic materials in this region.

Consequently, the $\delta^{13}C_{org}$ records are more similar to terrestrial records. In the eastern Tethys Ocean, particularly in the South China region, there is a greater prevalence of slope-deep basin facies sediments (Sun et al., 2016; Jin et al., 2018; Li et al., 2020). Concurrently, the sedimentation rates in the marine basins of South China diminished during the Carnian period (Dal Corso et al., 2024), which may have resulted in more pronounced alterations in $\delta^{13}C_{org}$ values due to the influx of terrestrial materials. This resulted in $\delta^{13}C_{org}$ records in the open platform or upper slope being more influenced by marine sediments, fluctuating primarily between -29‰ and -26‰ (Fig. 4, 5). Conversely, during the CPE, they were more influenced by terrestrial clastic materials, potentially fluctuating to around -24‰, which caused positive excursions in $\delta^{13}C_{org}$ and decoupled it from $\delta^{13}C_{carb}$ (Fig. 7A). In the lower slope or basin, sediment is less affected by terrestrial materials, resulting in $\delta^{13}C_{org}$ records that are more consistent with open ocean records (Fig. 7A).

## 6. Conclusions

This study conduct $\delta^{13}C_{carb}$ and $\delta^{13}C_{org}$ analysis on the Xiashulao section in the Lijiang Basin of South China, as well as the Erguan and Duanqiao sections in the Nanpanjiang Basin. The findings revealed a decoupling of $\delta^{13}C_{carb}$ and $\delta^{13}C_{org}$ during the Carnian period. Carbonate carbon isotope demonstrated a negative excursion of 2-3‰, whereas organic carbon isotopes exhibited a positive excursion of approximately 3-4‰. Low Mn/Sr ratio (<3) in most of the samples indicates that the $\delta^{13}C_{carb}$ curves recorded in the study sections are reliable. Positive shifts in $\delta^{13}C_{org}$ leads to the decoupling. During 1-2 CIE interval of CPE, C/N ratio raised to ~10 and Y/Ho decreased to 29 with high $\delta^{13}C_{org}$ values (~-25‰). It shows that the main factor for the positive shift in $\delta^{13}C_{org}$ due to increased terrestrial input. Changes in Sr/Ba ratios (from >100 to <1) link increased terrestrial input to the enhanced water cycle, providing further insights into decoupling of $\delta^{13}C_{carb}$ and $\delta^{13}C_{org}$. The sediments in the upper shelf and lower slope are more conducive to the influence of terrestrial and pelagic organic matter, respectively. Consequently, no decoupling of $\delta^{13}C_{carb}$ and $\delta^{13}C_{org}$ is observed in upper shelf and lower slope. In the lower shelf to upper slope regions, the composition of $\delta^{13}C_{org}$ is influenced by marine organic matter during background periods and by terrestrial materials during the CPE, resulting in the decoupling of $\delta^{13}C_{carb}$ and $\delta^{13}C_{org}$. Moreover, the current $\delta^{13}C_{carb}$ and $\delta^{13}C_{org}$ records during the CPE indicate that changes in the sedimentary environment and terrestrial input are the primary factors controlling the $\delta^{13}C_{org}$ composition. The discrepancy in the contributions of marine and terrestrial organic matter to sedimentary $\delta^{13}C_{org}$ results in fluctuations in carbon isotope records across diverse geographical regions.

## Acknowledgements

Shuxun Yuan and Xin Yang is thanked for the field help. Jing Li is thanked for the assistance during sample preparation for carbon isotope test. This research was supported by the State Key R&D Project of China (2023YFF0804000), and the National Natural Science Foundation of China (92155201, 92255303 and 42325202).

# Appendix A

**Geochemical analysis**

$\delta^{13}C_{carb}$-VPDB and $\delta^{18}O_{carb}$-VPDB

Weigh 400 μg of sample into a 10 mL borosilicate headspace vial and place it in a thermostatic reaction chamber, run the He Flush program for 750s to completely replace the air in the vial with 99.999% He, then add 0.05 mL of 99.9% phosphoric acid to the vial and react at 72°C for 1 h. The $CO_2$ generated by the reaction was mixed with He carrier gas and then passed through the water trap in the GasBench, quantitative valve and column, and finally sent to the mass spectrometer for testing. After mixing with the He carrier gas, the $CO_2$ generated from the reaction was passed through the water removal trap in the GasBench, the dosing valve and the chromatographic column, and finally sent to the mainframe mass spectrometer for testing. The analytical error of $\delta^{13}C$ is generally less than 0.1‰; the analytical error of $\delta^{18}O$ is generally less than 0.2‰. Reference standard: $\delta^{13}C$: GBW04416: 1.61‰、GBW04417: -6.06‰; $\delta^{18}O$: GBW04416: -11.59‰、GBW04417: -24.12‰.

$\delta^{13}C_{org}$-VPDB and TOC

Weigh 0.5-12 mg of HCl treated samples, wrapped into a 4* 6mm silver cup, seal the opening tightly, compacted into a small square, sent to the EA injection tray automatically in the carrier gas driven high-temperature oxidation, converted into $CO_2$, through the magnesium perchlorate absorption trap to remove $H_2O$, 70°C column separation of $N_2$ and other impurities, and ultimately sent to the main mass spectrometry test. The standard deviation of the current specimen is about

0.1‰. General sample error is not more than 0.2‰. Reference standard: USGS40: -26.39‰; USGS24: -16.05‰; IVA33802174 (Urea): -37.32‰.

TON

The sample was weighed 0.5-7 mg, wrapped tightly in a 5×9 mm aluminum foil cup, and placed into a Vario MACRO cube elemental analyzer using an autosampler. The sample was passed through the oxidation tube at 1150°C and the reduction tube at 850°C by the carrier gas successively to generate $N_2$, $CO_2$, $SO_2$ and $H_2O$. In the water trap, $H_2O$ was absorbed by $P_2O_5$, and $CO_2$ was absorbed and released from the $CO_2$ and $SO_2$ traps successively to separate $N_2$, $CO_2$ and $SO_2$, and then finally, the peaks of $N_2$, $CO_2$ and $SO_2$ were detected by the TCD detector successively.

Major element analyses

The sample pretreatment of whole rock major element analysis was made by melting method. The flux is a mixture of lithium tetraborate, lithium metaborate and lithium fluoride (45:10:5), Ammonium nitrate and lithium bromide were used as oxidant and release agent respectively，The melting temperature was 1050℃ and the melting time was 15min. The X-ray tube of X-ray fluorescence spectrometer (XRF) is a 4.0Kw end window Rh target, The test conditions are voltage: 50kV, current: 60mA.

Trace element analyses

The detailed sample-digesting procedure was as follows: (1) Sample powder (200 mesh) were placed in an oven at 105°C for drying of 12 hours; (2) 50mg sample powder was accurately weighed and placed in a Teflon bomb; (3) 1ml HNO$_3$ and 1ml HF were slowly added into the Teflon bomb; (4) Teflon bomb was putted in a stainless steel pressure jacket and heated to 190°C in an oven for >24 hours; (5) After cooling, the Teflon bomb was opened and placed on a hotplate at 140°C and evaporated to incipient dryness, and then 1ml HNO$_3$ was added and evaporated to dryness again; (6) 1ml of HNO$_3$, 1ml of MQ water and 1 ml internal standard solution of 1ppm In were added, and the Teflon bomb was resealed and placed in the oven at 190°C for >12 hours; (7) The final solution was transferred to a polyethylene bottle and diluted to 100g by the addition of 2% HNO$_3$.

**Biostratigraphy and Facies analysis**

Biostratigraphy analysis

In this study, we report for the first time the biostratigraphy of the Beiya and Zhongwo formations in the Lijiang Basin, and the Gaicha and Sanqiao formations in the Nanpanjiang Basin. In the Xiashulao section, two conodont zones were identified: the *Q. polygnathiformis* Zone and the *Q. polygnathiformis - noah* assemblage Zone (Fig. S1A). In the Erguan section, two conodont zones were identified from bottom to top: the *Quadralella polygnathiformis* Zone and the *Q. polygnathiformis - noah* assemblage Zone (Fig. S2B). In the Duanqiao section, one conodont zone was identified in the Zhuganpo Formation: *Q. polygnathiformis* Zone (Fig. S1C).

The *Q. polygnathiformis* Zone is defined by the first occurrence of *Quadralella polygnathiformis* as its base and the appearance of *Q. noah* as its top, indicating the boundary between the Ladinian and Carnian stages. The first appearance of *Q. polygnathiformis* in the Xiashulao section is found in the limestone at the top of the Beiya Formation (Fig. S2C), at a sampling position of 16 m. In the *Q. polygnathiformis* Zone at Xiashulao section, *Quadralella*

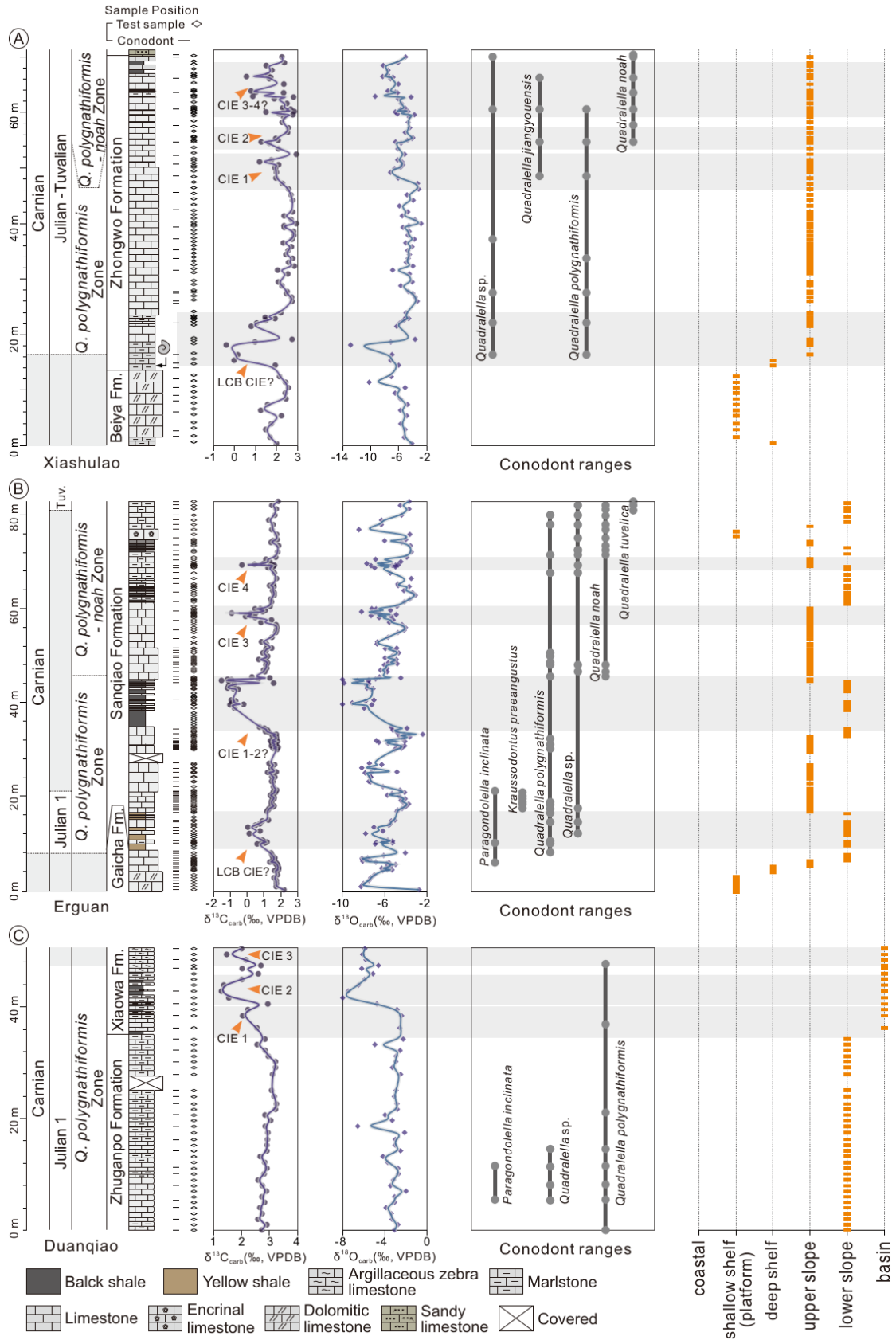

**Fig. S1.** Lithology, biostratigraphy, sample position, $\delta^{13}C_{carb}$, and $\delta^{18}O_{carb}$, distribution of conodonts and facies changes of studied sections. (A) Xiashulao section from Lijiang Basin. (B) Erguan section

from Nanpanjiang Basin. (C) Duanqiao section from Nanpanjiang Basin. Black arrows indicate the main intervals of terrestrial inputs and enhanced water circulation. Fm.: Formation. Tuv.: Tuvalian.

*jiangyouensis* and *Quadralella* sp. were also found. Additionally, *Xenoprotrachyceras* sp.(Fig. S2R1-4), a Ladinian ammonoid fossil of South China, was found in the limestone at the base of the Zhongwo Formation in the Xiashulao section at a sampling position of 14 m. At the Erguan section, the first appearance of *Q. polygnathiformis* is in the limestone at the top of the Gaicha Formation (Fig. S2D), at a sampling position of 8.1 m. In the *Q. polygnathiformis* Zone, the *Q. polygnathiformis are* with co-occurring species *Paragondolella inclinata* (Fig. S2A, B) and *Kraussodontus praeangustus* (Fig. S2I, J). Below this, at a sampling position of 6.21 m, the common species near the Ladinian-Carnian boundary, *P. inclinata*, was found. The Q. *polygnathiformis Zone* start from the bottom of Duanqiao section, with *Paragondolella inclinata* and *Quadralella* sp.

The *Q. polygnathiformis - noah* assemblage Zone has its base defined by the appearance of *Q. noah*, with no defined top. This assemblage Zone may indicate the upper of Julian and lower part of Tuvalian substages. In the Xiashulao section, *Q. noah* first appears at a sampling position of 54.3 m (Fig. S2M), with co-occurring species *Quadralella polygnathiformis* and *Quadralella jiangyouensis* (Fig. S1A). In the Erguan section, *Q. noah* first appears in the limestone after the first black shale interval of the Sanqiao Formation (Fig. S2K, L), at a position of 45.8 m, extending to the top of the section with co-occurring species *Quadralella polygnathiformis and Quadralella tuvalica* (Fig. S2N-Q).

In summary, in the Xiashulao River section, the transition of the Ladinian-Carnian boundary is located between 14 m and 15.8 m, and the Julian–Tuvalian boundary is tentatively in the *Q.*

*polygnathiformis - noah* assemblage Zone. In the Erguan section, the position below the first appearance of *Q. polygnathiformis* may be the transition of the Ladinian-Carnian boundary, and the Julian–Tuvalian boundary may be above the sampling position of 45.8 m and below the occurrence of *Quadralella tuvalica*.

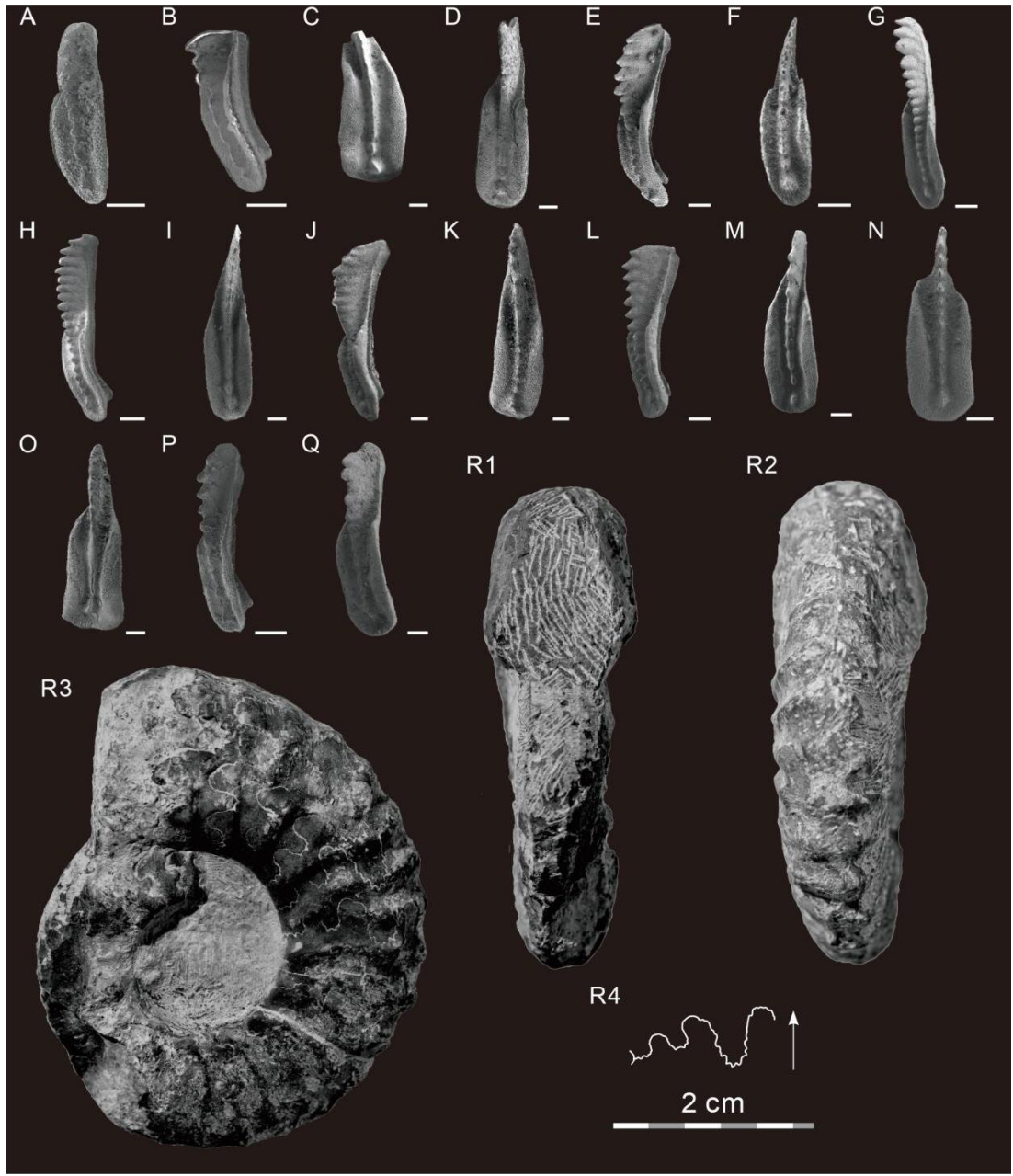

**Fig. S2.** Images of conodonts and ammonid from studied sections. Scale bars = 100 μm. A, B, *Paragondolella inclinata* Kovács, A, upper view, SSC-sq5-1; B, lateral view, SSC-sq9-1. C-H, *Quadralella polygnathiformis* (Budurov and Stefanov), C, D, F, upper view, XSLHB-49-1, SSCA-

9-1 and DQA-1-1; E, G, H, lateral view, SSC-sq14-1, DQA-1-2, DQA-1-3. I, J, *Kraussodontus praeangustus* (Kozur, Mirauta and Mock), I, upper view, SSC-sq20-1; J, lateral view, SSC-sq18-1. K-M, *Quadralella noah* Hayashi, K, M, upper view, EGA-sq1-1 and XSLHB-49; L, lateral view, EGC-sq112-1. N-Q, *Quadralella tuvalica* Mazza and Rigo, N, O, upper view, EGA-sq151-1 and EGA-sq151-2. R1-4, *Xenoprotracbyceras* sp.

Facies analyses

During the Middle to Late Triassic changes in facies were recorded in the Xiashulao, Erguan and Duanqiao section. Three facies type were identified from the upper part of the Beiya Formation to the Zhongwo Formation from Xiashulao section (Fig. S1A). At Xiashulao, the 14 m dolomitic limestone (Fig. S3A) in the upper part of the Beiya Formation corresponds to a tidal flat environment within the platform (Fig. S1A). The base of the overlying Zhongwo Formation from ~14 m to 23 m can be interpreted as an open platform to deep shelf (Fig. S1A), with the appearance of Brachiopods, bivalves, gastropods, echinoderms (Fig. S3B). At the middle and upper part of Zhongwo Formation, it begins to change to a deep-water slope environment below the wave base (Fig. S1A). There is less biological detritus, mainly dominated by brachiopods, bivalves and small foraminifera ((Fig. S3C), and typical photosynthetic organisms, such as calcareous algae, are not seen, indicating a deeper-water depositional environment.

Five main facies types were identified in the Gaicha and Sanqiao Formation from the Erguan section in the Nanpanjiang Basin (Fig. S1B), and four deep to shallow cycles are visible (Fig. S1B). The Gaicha Formation mainly records the transformation of dolomitic limestone to shale and marlstone, which corresponds to the rapid transition from a platform to a shelf environment (Fig.

S3D-F). Raw detrital grains dominated by echinoderm fragments, foraminifera, and filamentous bivalve fragments are seen in the limestone interbedded by shale at the base of the Sanqiao Formation, which are poorly sorted and indicative of a deeper-water, lower-energy, shelf environment. In the light grey thick bedded limestone above the yellow shale, more echinoderm fragments, gastropod fossils (Fig. S3G-H), and brachiopod fossils are seen, and the increase in benthic organisms may suggest shallower water depths, which is presumed to be a shallow slope environment, as no typical platform fossils, such as calcareous algae, are found. In the part with thin layered limestone interbedded within the shale, brachiopods are visible, and the grains are mainly composed of sub-angular well-sorted quartz clasts and some calcareous clasts, and pyrite and spongy bony needles can be observed (Fig. S3J), which suggest a deep water environment and the increase of the terrestrial inputs. In the middle and upper part of the Sanqiao Formation, the shale with interbedded limestone decreases gradually in proportion. Layers dominated by continuous greywacke deposits are then predominantly midland shelf environments (Fig. S3I). It is worth mentioning that a set of reef-limestone dominated by sea lilies (Fig. S3K) and corals were deposited at 75-77 m in Erguan. It is suggested that the occurrence of reef may corresponds to a terrace fringing reef environment.

The Zhuganpo and Xiaowa Formations in the Duanqiao section as a whole record a deeper-water shelf-basin depositional environment, with four main facies types identified (Fig. S1C). The Zhuganpo Formation consists of bioclastic marl-limestone. Thin-shelled bivalves (fine filaments) and spongy bony needles are abundant, and some small foraminifera are seen in a deep-water shelf environment (Fig. S3L, M). The overlying Xiaowa Formation consists mainly of black shale interbedded with thin layers of mud-stripped greywacke, in which radiolarians and filamentous bivalves are abundant, and, unlike the Erguan, no quartz fragments are seen (Fig. S3N, O). It indicate a deeper environment, possibly a basin.

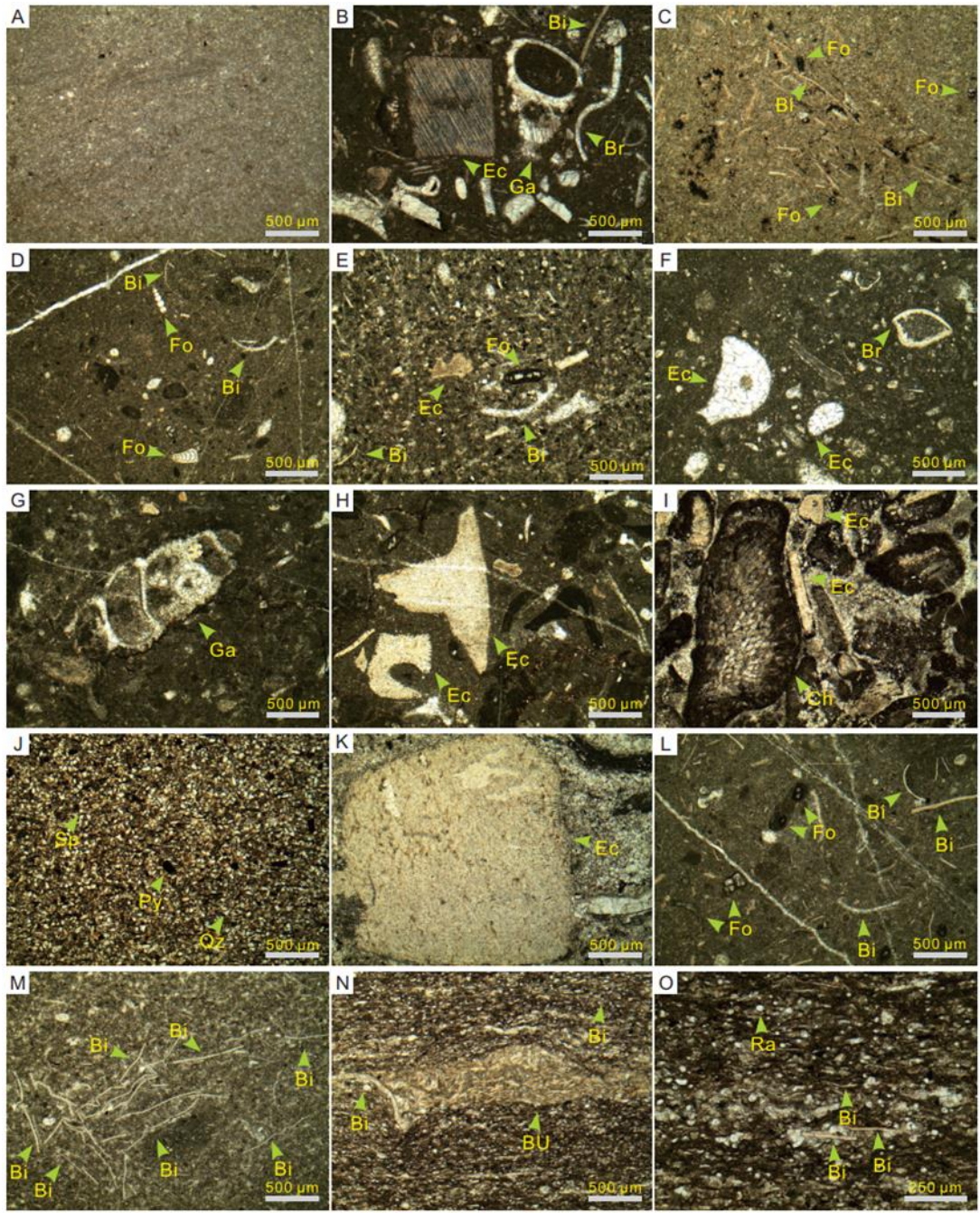

**Fig. S3.** Polarized microscope photographs of carbonate samples from Xiashulao (A-C), Erguan (D-K), and Duanqiao (L-O). (A) Non-laminated, non-biotic, dolomitic mudstone limestone, sample X2-10. Bioclastic mudstone limestone, with arrows indicating major bioclastic particles: brachiopods (Br), bivalves (Bi), gastropods (Ga), echinoderms (Ec), sample X2-93. Bioclastic

mudstone limestone, with arrows indicating major bioclastic particles as bivalves (Bi) and foraminifera (Fo), sample Bra-79. (D) Bioclastic marl-limestone, arrows indicate the main bioclastic particles: Foraminifera (Fo), Bivalves (Bi), sample SSCA-gc2c. (E) Bioclastic marl-limestone, arrows indicate the main bioclastic particles: Bivalves (Bi), Foraminifera (Fo), Echinoderms (Ec), sample EGC-146. (F) Bioclastic marl-limestone, arrows indicate the main bioclastic particles: Echinoderms (Ec), Brachiopods (Br), sample SSCA-77. (G) Bioclastic gainstone, arrows indicate the main bioclastic particles are Gastropods (Ga), with slight micritization visible at the margins of these particles, sample SSCB-13. (H) Bioclastic gainstone, arrows indicate the main bioclastic particles are Echinoderms (Ec), with these particles showing signs of abrasion, sample EGA-2h. (I) Limestone with mud and coated bioclastic particles, arrows indicate the main bioclastic particles are sponges (Ch) and Echinoderms (Ec), sample EGD-11. (J) Sandy marlstone, with subangular, well-sorted detrital quartz particles (Qz), and dark minerals being pyrite (Py), along with a small amount of sponge spicules (Sp), sample SSCB-48. K. Crinoidal bioclastic limestone, where the main particles are crinoidal bioclasts (Ec), sample EGC-99. (L) Bioclastic wackestone with arrows indicating major bioclasts: foraminifera (Fo), bivalves (Bi), ostracods (Os), sample DQA-41. (M) Thin-shelled pelagic bivalve wackestone, with grains primarily composed of thin and curved bivalve (Bi) shells, samples DQA-74. (N) Bioclastic-intraclastic silty limestone, with arrows indicating intraclasts composed mainly of bioclastic particles like bivalves (Ec) and disorganized fibrous bioclasts (BU) caused by bioturbation, sample DQA-171. (O) Radiolarian wackestone, with arrows indicating major bioclastic particles: bivalves (Bi) and recrystallized radiolarians (Ra?) with unpreserved internal structures, with bioclasts slightly aligned, sample DQA-172.